\documentclass[prb,aapm,mph,amsmath,amssymb,reprint,floatfix]{revtex4-1}

\usepackage{latexsym}
\usepackage{graphicx}

\usepackage{subfigure}    

\usepackage{verbatim}
\usepackage{amsmath}
\usepackage{color}
\usepackage{xcolor}
\usepackage{soul}

\usepackage{placeins}
\usepackage{amsmath}
\usepackage{amssymb}
\usepackage{amsthm}
\usepackage{amsfonts}
\usepackage{braket}

\usepackage{latexsym}
\usepackage{graphicx}

\usepackage{graphicx}
\usepackage{grffile}
\usepackage{hyperref}

\begin{document}

\title{Imaginary-time time-dependent density functional theory for periodic systems}

\author{John McFarland$^{(1)}$} 
\email{swqecs@gmail.com}
\author{Efstratios Manousakis$^{(1,2)}$}
\email{manousakis@gmail.com}
\affiliation{$^{(1)}$Department  of  Physics and National High Magnetic Field Laboratory, Florida  State  University,  Tallahassee,  FL  32306-4350,  USA}
\affiliation{$^{(2)}$Department   of    Physics,  National and Kapodistrian University    of   Athens,
 Panepistimioupolis, Zografos, 157 84 Athens, Greece}

\begin{abstract}
  Imaginary-time time-dependent Density functional theory (it-TDDFT) has been proposed as an alternative method for obtaining the ground state within density functional theory (DFT) which avoids some of the difficulties with convergence encountered by the self-consistent-field (SCF) iterative method.  It-TDDFT was previously applied to clusters of atoms where it was demonstrated to converge   in select cases where SCF had difficulty with convergence.  In the present work we implement it-TDDFT propagation for {\it periodic systems} by modifying the Quantum ESPRESSO package,
  which uses a plane-wave basis with multiple $\boldsymbol{k}$ points, and has the options of non-collinear and DFT+U calculations using ultra-soft or norm-conserving pseudo potentials.
  We demonstrate that our implementation of it-TDDFT propagation with multiple $\boldsymbol{k}$ points is correct for DFT+U non-collinear calculations and for DFT+U calculations with ultra-soft pseudo potentials.  Our implementation of
  it-TDDFT propagation converges to the exact SCF energy (up to the
  decimal guaranteed by double precision) in all but one case where it converged to a slightly lower value than SCF, suggesting a useful alternative for systems where SCF has difficulty to reach the Kohn-Sham ground state.
  In addition, we demonstrate that rapid convergence can be achieved if we use  adaptive-size imaginary-time-steps for different kinetic-energy plane-waves.   

\end{abstract}

\maketitle

\section{Introduction}

Having excellent scalability, density functional theory (DFT) is widely used for electronic structure calculations in a variety of fields.   DFT scales as $O(N^2)$ with number of electrons  up to around 1000 electrons, and beyond that it scales as $O(N^3)$.\cite{SCF} It is based on the Hohenberg-Kohn  theorem\cite{HohenbergKohn} of a one-to-one correspondence between the ground-state wavefunction and its charge density.  Methods that search for the ground state, seek the lowest energy solution where both the wavefunction and charge density are consistently calculated from each other.  

The Kohn-Sham (KS) Hamiltonian is a single particle Hamiltonian with a potential that is the sum of the coulomb potential from the atomic cores and the electron charge-density plus the exchange-correlation potential.\cite{kohnSham} The latter is a universal functional of charge density and it is usually tuned to give the same energy-density dependence as that of an interacting-homogeneous electron-gas.\cite{CeperlyAdler,GGA}  

The KS wave function is a set of single particle wave functions (KS states) that are occupied by one or two electrons for the collinear and non-collinear cases respectively.  When in the ground state, the KS states are the lowest-energy eigenvectors of the KS Hamiltonian.  The KS ground state and Hamiltonian are required to be consistent with each other.

The KS ground state is typically reached by means of self-consistent-field (SCF) iterations.\cite{SCF}  With an initial charge density, a set of lowest eigenvectors of the KS Hamiltonian are first calculated, then occupation numbers (weights) are assigned to them, and then a new charge density is calculated from the occupied eigenvectors.  This density is usually mixed with densities from previous iterations to aid convergence.\cite{PULAY1980393}

SCF has difficulty converging for some systems.  One reason is charge sloshing, where some orbitals continually fluctuate above and below the Fermi level causing their weights, and, thus, the charge density to fluctuate as well.  This is more prevalent for systems with many bands close to the Fermi level, such as metallic and large systems.\cite{convergence}  Mixing a smaller fraction of the calculated charge density with previous iterations and the smearing of occupation numbers close to the Fermi level are common solutions.\cite{smearing}

Recently imaginary-time time-dependent Density functional theory (it-TDDFT) propagation was proposed by Flamant {\it et. al.}\cite{harvard} as a more reliable method for generating the KS ground state in various cases of atomic clusters where the SCF iteration approach has difficulty converging to the KS ground-state.
In this approach the KS orbitals are propagated in imaginary time
with the KS Hamiltonian, while their orthonormalization is maintained.  
By extending a proof by Van Leeuwen \cite{VanLeeuwen} to it-TDDFT propagation, Flamant {\it et. al.} suggest that it-TDDFT propagation will eventually converge to the KS ground-state, provided that the initial state has a non-zero overlap
with the KS ground-state, and that a sufficiently small time-step is used which they adjust ``on-the-fly''. They demonstrate successful convergence to the ground states of $\mbox{Cu}_{15}$ and $\mbox{Ru}_{55}$ nanoclusters,  where various SCF iteration schemes
either converge to a higher energy state or fail to converge at all. They used code based on SIESTA,\cite{Siesta} which uses a localized basis.

In  the present  work  we  have applied  it-TDDFT  propagation to
extensive (periodic)
systems  by  modifying  the   open  source  package  Quantum  ESPRESSO
(QE),\cite{Giannozzi_2009,Giannozzi_2017}   which    models   periodic
systems using a  plane-wave basis.  For periodic  systems, KS orbitals
  take the  Bloch form  and essentially  gain three  additional
dimensions   corresponding    to   the   values   of    the   momentum
$\boldsymbol{k}$ in the Brillouin zone.
The charge  density is  calculated by
integration of  occupied  band charge  densities within  the first
Brillouin  zone.  

First,  we demonstrate  that for  simple systems  our
implementation gives the  same ground-state energy and  density as 
SCF up  to 13  or more  decimal places. Second,  we choose  a metallic
system (FCC  copper) and demonstrate  that we can reproduce  the same
ground-state energy, charge density, and Fermi  surface as
SCF.  We also demonstrate that it-ITDDFT can find a state with lower energy than SCF.
Our     implementation    can     be     found     at
\url{https://github.com/Walkerqmc/ITDDFT_for_QE})  and, therefore, is now part of the public domain to be applied in cases where SCF loops appear
to have difficulty reaching a fully converged ground state or
when it is desirable to validate by a very different method that SCF
iterations have reached
the true DFT ground state.

The paper is organized as follows.  In Sec.~\ref{implementation}
we discuss our formulation of the problem and how it has been implemented
using the quantum espresso. In Sec.~\ref{results} we apply the method
on selected systems and demonstrate that our implementation of
it-TDDFT propagation for periodic systems reproduces SCF results for the
ground-state energy, density, and Fermi surface, except in one case where it finds a lower energy state. In Sec.~\ref{conclusions} we present our main conclusions.
   
\section{Imaginary Time Propagation and Implementation for Quantum ESPRESSO}
\label{implementation}

\label{it-TDDFT-propagation}
The ground-state wavefunction within DFT consists of single-particle
orbitals $ \ket {\psi_m}$.  Assuming their orthonormalization, i.e., $ \braket{\psi_j|\psi_m}=\delta_{jm}$, the charge density $\rho$ is given by 
\begin{equation}  
\label{eqn:rho}
\rho(\textbf{r})  =  \sum_{m\in occ.}  \braket{ \psi_{m} | \boldsymbol{r} } \braket{ \boldsymbol{r} | \psi_{m} },
\end{equation}
where the summation is over occupied orbitals.
The KS Hamiltonian is a function of $\rho$ and can be used to propagate
the orbitals in imaginary time $\tau=it$ according to the differential equation\cite{harvard}     
\begin{equation}  
\frac{d\ket {\psi_{m}}}{d\tau} = -{\mbox{H}}[\rho] \ket {\psi_{m}}.
\end{equation}
This evolution in it-TDDFT can be done efficiently with the forward Euler propagator
\begin{equation}  
\label{eqn:Euler}
\ket {\psi_{m}(\tau+\Delta\tau)} = \big(\mbox{I}-{\Delta\tau\mbox{H}}[\rho(\tau)] \big)\ket {\psi_{m}(\tau)}.
\end{equation}
Orthonormalization is not preserved with it-TDDFT propagation, and if we wish to calculate $\rho$ with Eq.~\ref{eqn:rho} we must first orthonormalize the
Bloch wavefunctions.  We have used the Gram-Schmidt method which successively orthonormalizes a set of orbitals, where if orbitals with index less than $m$ are orthonormal, the orbital $\ket {\psi_{m}(t)}$ is orthonormalized by the transformations
\begin{equation}  
\label{eqn:gramshmidt1} 
\ket {\psi_{m}} \Rightarrow \ket {\psi_{m}}  - \sum_{n<m} \ket {\psi_{n}} \braket{\psi_{n}|\psi_{m}},
\end{equation}
\begin{equation}  
\label{eqn:gramshmidt2}  
\ket {\psi_{m}} \Rightarrow \frac{\ket {\psi_{m}}}{\sqrt{  \braket{\psi_{n}|\psi_{m}}  }  }.
\end{equation}


Quantum ESPRESSO (QE) is used to solve the KS equations for periodic systems.  The KS orbitals are the Bloch states  $ \psi_{n\boldsymbol{k}}(\boldsymbol{r})=e^{-i\boldsymbol{k}\cdot\boldsymbol{r}}\large{u}_{n\boldsymbol{k}}(\boldsymbol{r})$ where
$\large{u}$ is periodic and expanded in  a plane-wave basis.
The index ${\bf k}$ forms  grid points within the first Brillouin zone.
Bloch states at different  $\boldsymbol{k}$ are characterized by
different kinetic energy of the plane-waves, allowing an effective $\boldsymbol{k}$-dependent Hamiltonian which QE uses to solve the eigenvalue equation
\begin{equation}  
\label{eqn:eigen}
\mbox{H} \ket {\psi_{n\boldsymbol{k}}}=\epsilon_{n\boldsymbol{k}}\ket {\psi_{n\boldsymbol{k}}}
\end{equation}  
separately for each value of $\boldsymbol{k}$ during SCF iterations.  

The next step of the iteration is to calculate the charge density and total energy, and to do this occupation numbers called weights $w_{n\boldsymbol{k}}$ are assigned to the Block wavefunctions.  The weights are normalized so they yield
the correct total charge of the valence electrons.  Generally, the bands up to Fermi level are occupied, with possible smearing near the transition from occupied to unoccupied states in order to aid convergence.
The charge density and band energy are then calculated as follows
\begin{equation}  
\label{eqn:rho2}
\rho(\textbf{r})  =  \sum_{n\boldsymbol{k}}  w_{n\boldsymbol{k}}\braket{ \psi_{n\boldsymbol{k}} | \boldsymbol{r} } \braket{ \boldsymbol{r} | \psi_{n\boldsymbol{k}} },
\end{equation}   
\begin{equation}  
\label{eqn:bandenergy}
E_{band} =\bra{\Psi}\mbox{H} \ket{\Psi} = \sum_{n\boldsymbol{k}}  w_{n\boldsymbol{k}}\epsilon_{n\boldsymbol{k}}.
\end{equation}  



Most $\boldsymbol{k}$ points can be grouped into sets related by symmetry transformations that preserve charge density and energy.  For efficiency, QE reduces the $\boldsymbol{k}$ points in such a set to a single one with a larger weight.  After the charge density is calculated, it is typically mixed with a linear combination of charge densities from previous iterations to aid convergence.

We implemented it-TDDFT by modifying the SCF iterations in the QE package.
Instead of solving the eigenvalue Eq.~\ref{eqn:eigen}, we apply the forward Euler propagator given by Eq.~\ref{eqn:Euler} to each of the Bloch states.  Gram-Schmidt orthonormalization is then performed separately for each group of Bloch states that have the same $\boldsymbol{k}$ index, since states with different $\boldsymbol{k}$ remain orthogonal over all cells after propagation.  We calculate the band energy as

\begin{equation}  
\label{eqn:states}
 \epsilon_{n\boldsymbol{k}} = \bra {\psi_{n\boldsymbol{k}}} \mbox{H} \ket {\psi_{n\boldsymbol{k}}},
\end{equation}
 
The weights of Eq.~\ref{eqn:bandenergy} must then be assigned to the Bloch states, and the method to do so must allow the transition of electrons to different $\boldsymbol{k}$-points if the ground state has a different $\boldsymbol{k}$-point distribution from the initial state.  We experimented with two weight-assigning methods.  One was to simply let one of the QE subroutines assign weights based on the $\epsilon_{n\boldsymbol{k}}$ values.  This, however, does not yield proper imaginary-time dynamics, since electrons can immediately transition from one Bloch state to another if the states move above and below the Fermi level during the it-TDDFT propagation. 

Since accurate imaginary-time dynamics may converge better by avoiding charge sloshing, does not require the use of smearing, and since an accurate history of the system energy could also be used to calculate the density of excited states, we tried another method that assigns weights in a way that approximates imaginary-time dynamics.  
We do this by dropping the requirement that a KS state is a Bloch state, since it-ITDDFT propagation does not change the $\boldsymbol{k}$ value of a Bloch state.  Instead we superimpose the KS states over Bloch states, using the latter as a basis.  In this case propagation with Eq.~\ref{eqn:Euler} will move the charge of a KS state from higher energy Bloch states to lower energy ones. Thus by initially superimposing the KS states among all possible $\boldsymbol{k}$-points, we allow it-TDDFT propagation to shift the electron $\boldsymbol{k}$-point distribution in a way that lowers energy.

We implement this by introducing a new set of coefficients  $c_{in\boldsymbol{k}}$ that relate the Bloch states in QE $\ket{ {\psi_{n\boldsymbol{k}}} }$ to the KS states $\ket {\phi_i}$, 
\begin{equation}  
\label{eqn:states}
\ket {\phi_i} = \sum_{n\boldsymbol{k}} c_{in\boldsymbol{k}}   \ket{ {\psi_{n\boldsymbol{k}}} }.
\end{equation} 
We initialize the $c_{in\boldsymbol{k}}$ values by first selecting them at random, then we Grahm-Schmidt orthonormalize them so that $ \braket{\phi_i|\phi_j}=\delta_{ij}$.   The KS states are then it-TDDFT propagated as
\begin{equation}  
\label{eqn:Euler2}
\ket {\phi_{i}(\tau+\Delta\tau)} = \sum_{n\boldsymbol{k}} c_{in\boldsymbol{k}}  \big(\mbox{I}-{\Delta\tau\mbox{H}}[\rho(\tau)] \big)\ket {\psi_{n\boldsymbol{k}}(\tau)},
\end{equation}
and then orthonormalized.  

To orthonormalize, we first give the KS states an orthonormal basis by orthonormalizing the $\ket{ {\psi_{n\boldsymbol{k}}}}$ states while simultaneously keeping the $\ket {\phi_i}$ states fixed by adjusting the $c_{in\boldsymbol{k}}$ coefficients.  This is done by building a $\boldsymbol{k}$-point dependent matrix during orthonormalization which later operates on the $c_{in\boldsymbol{k}}$ coefficients. After this we Gram-Schmidt orthonormalize the $c_{in\boldsymbol{k}}$ coefficients, and thus the KS states.

To calculate the charge density, we have employed an approximation that becomes exact as the wavefunction approaches the exact KS ground state.  Combining index $n$ and $\boldsymbol{k}$ to an index $m$, we calculate the weights as 
\begin{equation}  
\label{eqn:weights2}
w_{m} = \sum_{i} |c_{im}|^2,
\end{equation}  
and use Eqs.~\ref{eqn:rho2} and \ref{eqn:bandenergy} to calculate charge density and band energy.  The correct charge density is given by 
\begin{equation}  
\rho(\boldsymbol{r}) = \sum_{i}   \braket{ \phi_i | \boldsymbol{r} } \braket{ \boldsymbol{r} | \phi_i },
\end{equation}
which by mere substitution becomes
\begin{eqnarray}  
\rho(\boldsymbol{r}) &=& \sum_{i} \sum_{m'} \sum_{m} c^*_{im'} c_{im}  \braket{ {\psi_{m'}} | \boldsymbol{r} } \braket{ \boldsymbol{r} | \psi_{m} }.
\end{eqnarray} 
When approaching the KS ground state, the number of both our KS states and occupied Block states become equal.  This makes $c_{im}$ a unitary matrix resulting in $\sum_{i} \ c^*_{im'} c_{im} = \delta_{m,m'}$, and converges to our approximation for the charge density 
\begin{eqnarray}  
\rho(\boldsymbol{r}) &=& \sum_{i} \sum_{m} |c_{im}|^2   \braket{ \psi_{m} | \boldsymbol{r} } \braket{ \boldsymbol{r} | \psi_{m} }.	
\end{eqnarray} 

As explained, QE reduces the bands of a set of $\boldsymbol{k}$ points related by symmetry to one $\boldsymbol{k}$ point with accumulated weight $w_{\boldsymbol{k}}$, where $w_{\boldsymbol{k}}$ is the weight of a fully occupied Bloch state.  To account for this extra weight we assign additional $c_{in\boldsymbol{k}}$ coefficients to the bands of a reduced $\boldsymbol{k}$ point, equal to the $w_{\boldsymbol{k}}$ of that point divided by the greatest common factor of all $w_{\boldsymbol{k}}$ values.  We also reduced the number of $c_{in\boldsymbol{k}}$ coefficients by freezing the electrons well below the Fermi surface.

We implemented it-TDDFT for both norm-conserving\cite{NCPPs} and ultra-soft\cite{USPPs} pseudo-potentials (USPPs).  The formalism presented so far was for the former.  USPPs let us use less plane-waves (i.e., a lower energy cut-off) by smoothing out the wavefunction within cutoff radii of the nuclei.  The price that we pay for doing this is that the Bloch states are not orthonormal, but instead obey the condition 
\begin{equation}
\bra{ {\psi_{n\boldsymbol{k}}} } \mbox{S} \ket{ {\psi_{n'\boldsymbol{k'}}} } = \delta_{n\boldsymbol{k},n'\boldsymbol{k'}}.
\end{equation} 
$\mbox{S}$ can be thought of as mapping the bands $\ket{\psi_{n\boldsymbol{k}}}$ to a set of orthonormal bands $\ket{\tilde{\psi}_{n\boldsymbol{k}}}$, with $\ket{\tilde{\psi}_{n\boldsymbol{k}}} = \mbox{S}^{\frac{1}{2}}\ket{\psi_{n\boldsymbol{k}}}$.  

We can obtain the equations for the bands of USPPs from the above equations for orthonormal bands with the substitutions $\ket{\psi_{n\boldsymbol{k}}} \rightarrow \mbox{S}^{\frac{1}{2}}\ket{\psi_{n\boldsymbol{k}}}$ and $\mbox{H} \rightarrow \mbox{S}^{-\frac{1}{2}} \mbox{H} \mbox{S}^{-\frac{1}{2}}$, provided $\mbox{S}$ is constant.  As an example, with substitutions and multiplying the left-hand-side of Eq.~\ref{eqn:Euler} by $\mbox{S}^{-\frac{1}{2}}$, the forward Euler propagator becomes 
\begin{equation}  
\label{eqn:Euler3}
\ket {\psi_{m}(\tau+\Delta\tau)} = \big(\mbox{I}-{\Delta\tau\mbox{S}^{-1}\mbox{H}}[\rho(\tau)] \big)\ket {\psi_{m}(\tau)}.
\end{equation}
We implement Eq.~\ref{eqn:Euler3} by operating on the bands first with a Hamiltonian subroutine and then an inverse S subroutine found in the QE package.  The latter does not support non-collinear calculations, which is where plane-waves and spin are combined to a single basis to and enable relativistic spin orbit coupling.\cite{non-col}  Thus, we have only implemented it-TDDFT with USPPs for the collinear case.

Our implementation of it-TDDFT naturally works for DFT+U calculations, which model the effect of correlations by introducing an extra energy term U when two electrons occupy the same atomic site.\cite{DFT+U}  In this case the Hamiltonian depends also on occupation numbers $n_j$, calculated by 
\begin{equation}  
\label{eqn:occupation}
n_j = \sum_{n\boldsymbol{k}} \braket{\psi_{n\boldsymbol{k}} | \chi_j}\braket{ \chi_j | \psi_{n\boldsymbol{k}} },
\end{equation}
where $\ket{\chi_j}$ are local orbitals of atoms that a Hubbard U was applied to.

The size of the time-step is an important parameter in the it-ITDDFT propagation.  Generally one wants the largest value that still results in monotonically decreasing energy.  Flamant {\it et. al.}\cite{harvard} adjusted the time-step on the fly by decreasing it when the total energy increased and increasing it otherwise.  Our experience with this variable time-step was that in some cases it would constantly decrease to effectively zero, never resulting in monotonically decreasing energy.  We instead set the time-step equal to slightly less than two divided by the maximum plane-wave kinetic energy.  Since if the time-step is larger than this, the propagator of Eq.~\ref{eqn:Euler} will multiply the highest kinetic-energy plane-waves with a value less than negative one, resulting in exponential increase.

We have also developed a method that uses a larger time-step for lower energy plane-waves.  This propagation alone would result in disproportionately larger expansion of the lower-energy plane-waves, so we counter it by dividing each plane wave by the expected change in its magnitude given the  associated time step and band energy.  Specifically, we use the propagator
\begin{equation}  
\label{eqn:accprop1}
\ket {\psi_{i}} \Rightarrow \small{ \sum_{g} \ket{g} \bra{g}\frac{\mbox{I}-{\Delta\tau_g   \mbox{H}} }{1-\Delta\tau_g \epsilon_i}}\ket {\psi_{i}},
\end{equation}
where the states $\ket{g}$ stand for the plane-wave, and where
\begin{equation}  
\label{eqn:bandenergy2}
\epsilon_i = \bra{\psi_{i}}  \mbox{H}\ket{\psi_{i}}.
\end{equation}
Namely, as the bands approach eigenvectors of $\mbox{H}$, the propagator of Eq.~\ref{eqn:accprop1} approaches normalized imaginary-time propagation.  We set the plane-wave dependent time-step to the inverse of the plane-wave kinetic energy with a maximum cut-off.  Orthonormalization and the calculation of weights are performed in the same way as already discussed.


\section{Results}
\label{results}
\subsection{Benchmarking our implementation}
We present the results of the application of our implementation of
it-TDDFT within QE on various systems, demonstrating that it
can converge to the same  energy as SCF within the desired level of
accuracy.  In addition we measured the difference between charge densities of it-TDDFT and SCF produced ground states for the systems studied using the metric  
\begin{equation}  
\label{eqn:RhoDifference}
D[n,n_0] = \frac{1}{2} \int |n(\boldsymbol{r}) - n_0(\boldsymbol{r})|d^3 \boldsymbol{r},
\end{equation}
where $\int n(\boldsymbol{r})d^3 \boldsymbol{r}$ equals the number of electrons.

\begin{figure}
	\centering
	\includegraphics[width=1\linewidth]{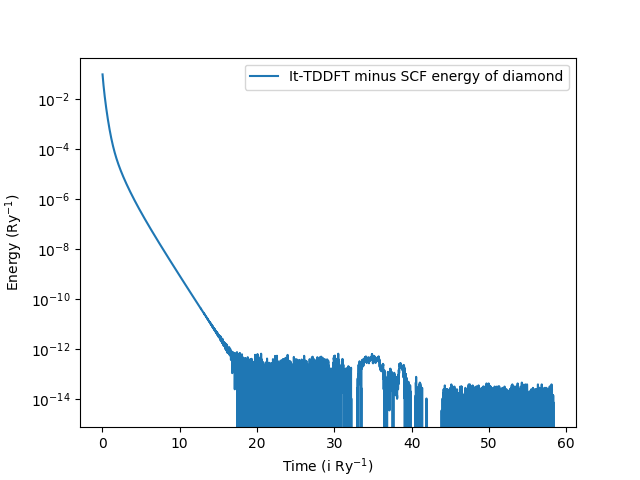}
	\caption{Energy of it-TDDFT propagated diamond minus SCF energy.}
	\label{fig:Fig1}
\end{figure}

\begin{figure}
	\centering
	
	\includegraphics[width=1\linewidth]{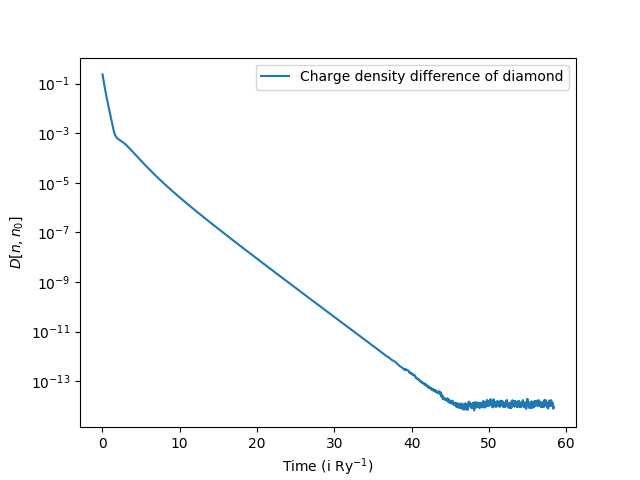}
	\caption{Charge-density difference between it-TDDFT propagated diamond and the SCF ground state, as defined by Eq.~\ref{eqn:RhoDifference},}
	\label{fig:Fig2}
\end{figure}

In Table~\ref{tab:5/tc} we compare the results of the application of
it-TDDFT on Si, Diamond, and Graphene to those obtained using
SCF iterations. Notice that the energy and the charge density $n({\bf r})$
in both calculations agree to at least 13 decimal places.
This is a strong indication that our implementation of it-TDDFT within QE
is correct.  We plot the energy and charge density difference as a function of imaginary-time in Fig.~\ref{fig:Fig1} and Fig.~\ref{fig:Fig2} respectively. Notice that the energy and density converge exponentially
and quickly to their SCF values up to 13-decimal points of accuracy and that
their difference from the SCF values falls within the numerical noise. 
\begin{table}[h]
	\centering
	\resizebox{\columnwidth}{!}{%
	\begin{tabular}{|l|c|c|c|}
		\hline
		System & $D[n,n_0]$ & SCF Energy ($Ry$) & it-TDDFT Energy ($Ry$)  \\
		\hline
		Silicon & $1.52\times10^{-15}$ & $-15.8535716860612$ & $-15.8535716860612$ \\
		\hline
		Diamond & $1.09\times10^{-14}$ & $-22.7713966524047$ & $-22.7713966524047$ \\
		\hline
	    Graphene & $1.49\times10^{-13}$ &$-22.791839022835$ & $-22.79183902283$ \\
		\hline
	\end{tabular}%
    }
	\caption{\label{tab:5/tc}Charge-density difference as defined by Eq.~\ref{eqn:RhoDifference} (second column) and ground state energy obtained with SCF
          (third column) and  it-TDDFT (fourth column) for three systems.  The ground state energy is in agreement to 13 or more decimal places.  Calculations were collinear with ultra-soft pseudo-potentials.}
\end{table}

\begin{figure}[h]
	\centering
	\includegraphics[width=1\linewidth]{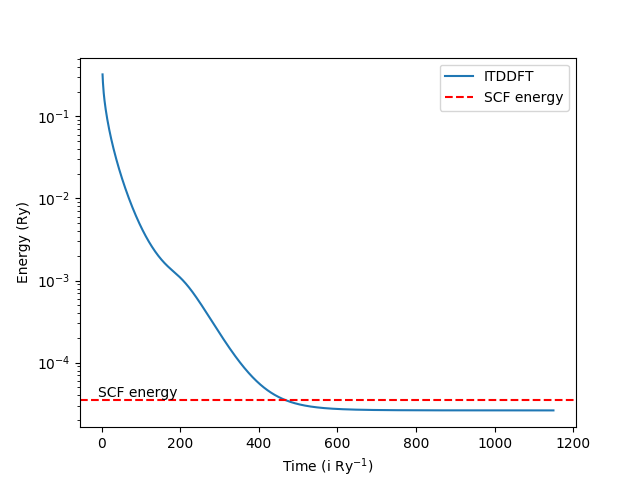}
	\caption{Relative energies of FCC copper it-TDDFT propagated with a $2.73\times10^{-2}$ Ry$^{-1}$ time-step. Occupations are calculated with a QE method.  This calculation uses norm-conserving pseudo-potentials, 63 reduced k-points, and a 2 eV rotationally invariant Hubbard U implementation of Liechtenstein et. al \cite{DFT+U2}.}
	\label{fig:Fig3}
\end{figure}

\begin{figure}[!htb]
	\centering
	\includegraphics[width=1\linewidth]{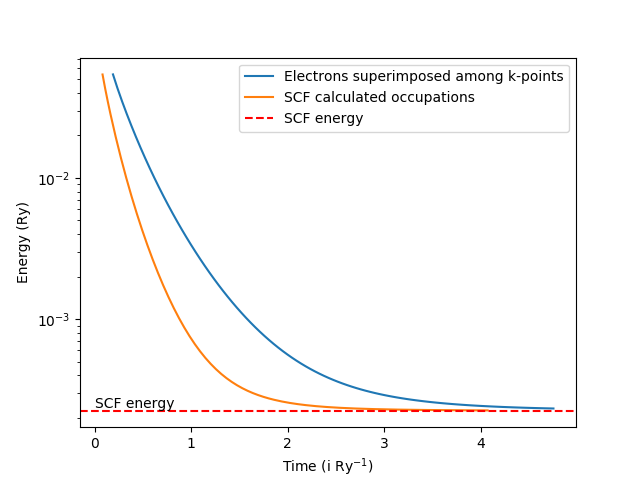}
	\caption{Relative energy of GaAs it-TDDFT propagated with a $2.73\times10^{-2}$ Ry$^{-1}$ time-step. For blue, electrons are superimposed among k-points.  For orange, occupations are calculated with a QE method.  Calculation uses a Hubbard U of 2 eV,  ultra-soft-pseudo potentials, and 8 reduced k-points.}
	\label{fig:Fig4}
\end{figure}

\begin{figure}
	\centering
	\includegraphics[width=1\linewidth]{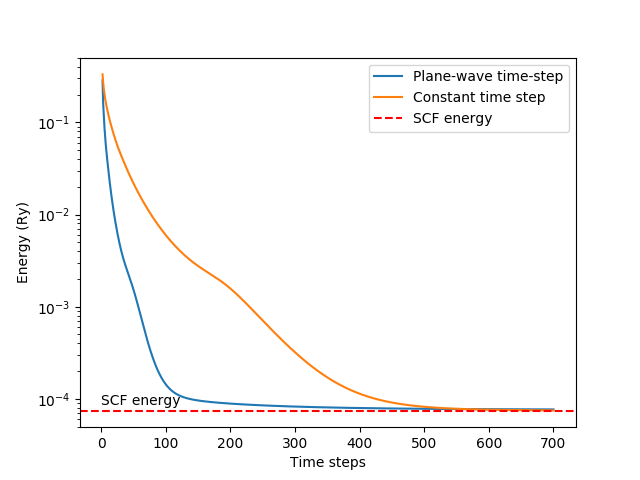}
	\caption{Relative energy of it-TDDFT propagated FCC copper.  Orange has a constant time-step equal to 1.9 divided by the maximum plane-wave kinetic energy, or $2.73\times10^{-2}$ Ry$^{-1}$.  Blue has a plane-wave dependent time-step equal to 1 divided by the plane-wave kinetic energy with a maximum cut off.  The calculation is collinear with no Hubbard U.}
	\label{fig:Fig5}
\end{figure}

\begin{figure}
	\centering
	\includegraphics[width=1\linewidth]{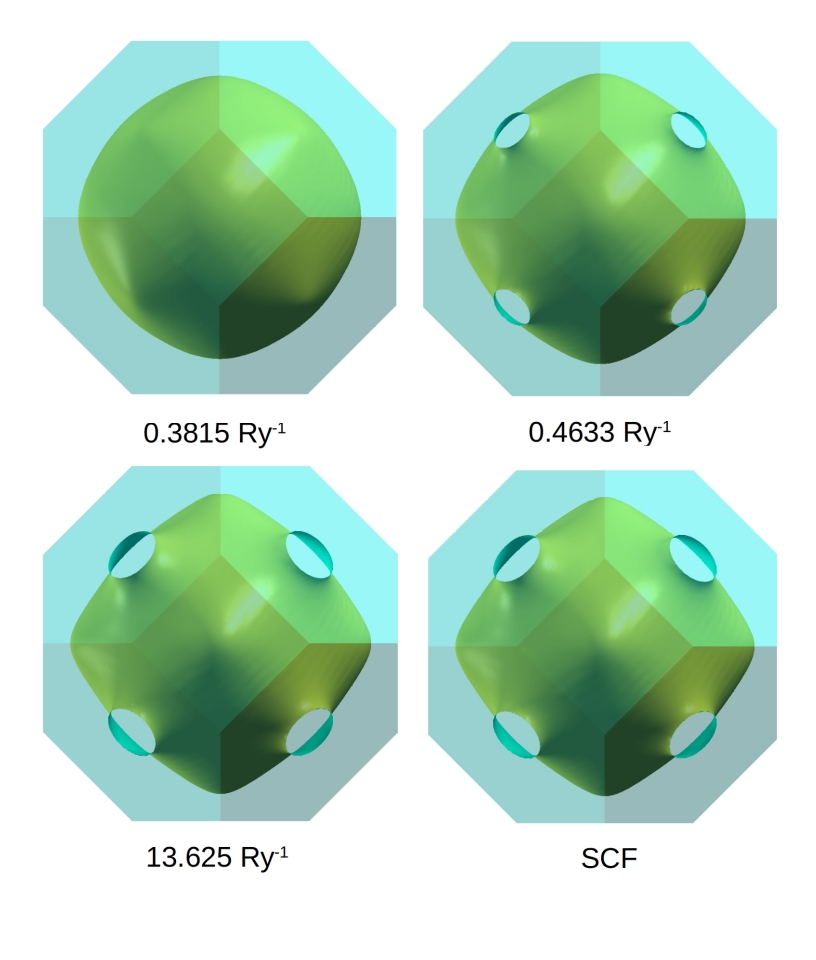}
	\caption{Imaginary time evolution of the Fermi surface of copper compared to that of SCF.}
	\label{fig:Fig6}
\end{figure}

\subsection{It-TDDFT can be advantageous to SCF}

In Fig.~\ref{fig:Fig3} we show that when adding a Hubbard U of 2 ev and performing a non-collinear calculation on FCC Copper, it-TDDFT propagation converges to a slightly lower energy of $-365.08915 (Ry)$, as compared to $-365.08914 (Ry)$ produced by SCF. The difference in it-TDDFT and SCF charge densities as defined by Eq.~\ref{eqn:RhoDifference} was $8.67\times10^{-6} (Ry)$.  SCF and it-ITDDFT calculations used the default settings of QE with norm-conserving pseudo-potentials, a $0.01$ value of the Gaussian spreading (smearing), a $0.3$ mixing factor for SCF, and a constant time step for it-TDDFT.  

Other cases where it-ITDDFT converges to a lower energy than SCF, indicating that the latter has difficulty
converging to the correct KS ground state, were presented in Ref.~\onlinecite{harvard}. Such cases are a reason one might find our application  of the it-ITDDFT method useful, namely, when SCF has difficulty to converge.  

\subsection{Additions to the method}

Next, in Fig.~\ref{fig:Fig4} we compare two methods of implementing it-TDDFT.  One using
superposition of Bloch states, i.e., using  Eq.~\ref{eqn:states}, to calculate occupations, the other using the standard QE method.  The system used for comparison is collinear GaAs
with a Hubbard U of 2eV using ultra-soft pseudo potentials.  Notice in
Fig.~\ref{fig:Fig4} that the superimposed Bloch states converge to exactly the same energy as SCF.  The speed of convergence of the superposition method is slower as compared to SCF. This can be explained by the small energy difference between different k-point distributions of the electrons, resulting in a slow shift of charge between k-points (whereas SCF calculated occupations can instantly shift charge).

While the superposition method is slower, 
it does not require smearing and could be useful when trying to converge to the correct Fermi surface if occupations near the Fermi energy shift from iteration to iteration due to small energy differences, i.e., charge sloshing. Also, for the copper system of Fig.~\ref{fig:Fig3}, the superposition method reached an energy that was $5\times10^{-5} (Ry)$ less than the standard QE method for calculating weights.

Next, in Fig.~\ref{fig:Fig5} we compare the speeds of the plane-wave dependent time-step with a constant time step for FCC copper.  Both converge close to the SCF ground state, with the plane-wave dependent time-step converging faster.  However, for some systems the plane-wave dependent time step fails to converge.






\FloatBarrier


\raggedbottom
\flushbottom
\section{Conclusions and discussion}
\label{conclusions}
Imaginary time propagation is an alternative method for calculating the DFT ground state, and may be useful for large metallic systems or any others where the standard SCF has difficulty converging to the correct ground state.\cite{harvard}  We have implemented it for periodic systems by modifying the open source package Quantum ESPRESSO.  Our implementation uses a plane-wave basis with the options of multiple $\boldsymbol{k}$-points, DFT+U, collinear or non-collinear, and norm-conserving or ultra-soft pseudo-potentials.  As we demonstrate, our implementation reproduces within the computer accuracy the results for the
ground-state energy and density distribution
obtained by SCF. We also demonstrate that the present method can
approach closer to the KS ground-state than the SCF in certain systems.
Therefore, it is an alternative to the SCF iterative scheme to
be at the disposal of the researcher who is interested in properties of periodic
systems.

We presented three variations of this implementation:
a) one that calculates the occupation numbers of the KS states using a Quantum
Espresso subroutine, b) one that calculates the occupation numbers by modeling the evolution of KS states that are superimposed over $\boldsymbol{k}$ points,
and c) another implementation that propagates lower kinetic energy plane-waves with a larger time-step.  We demonstrate the advantages and disadvantages of
these three implementations by applying them to several systems.

The source code of our     implementation    can     be     found     at
\url{https://github.com/Walkerqmc/ITDDFT_for_QE})  and, therefore,  it
can be easily applied in cases where the standard methodology of SCF
cannot reach a fully converged ground state or for other reasons, such as,
when it is desirable to make sure that SCF
iterations have reached the true DFT ground state.

\end{document}